\documentclass[12pt]{article}
\usepackage{myart}

\input tcilatex
\begin{document}

\title{Interfusion, labeling and stream lines in flow of inviscid barotropic
fluid }
\author{Yuri A.Rylov}
\date{Institute for Problems in Mechanics, Russian Academy of Sciences,\\
101-1, Vernadskii Ave., Moscow, 119526, Russia.\\
e-mail: rylov@ipmnet.ru\\
Web site: {$http://rsfq1.physics.sunysb.edu/\symbol{126}rylov/yrylov.htm$}\\
or mirror Web site: {$http://gasdyn-ipm.ipmnet.ru/\symbol{126}%
rylov/yrylov.htm$}}
\maketitle

\begin{abstract}
It is shown that the Euler system of hydrodynamic equations for inviscid
barotropic fluid for density and velocity is not a complete system of
dynamic equations for the inviscicd barotropic fluid. It is only a closed
subsystem of four dynamic equation. The complete system of dynamic equation
consists of seven dynamic equations for seven dependent variables: density,
velocity and labeling (Lagrangian coordinates, considered as dependent
variables). Solution of the Cauchy problem for the Euler subsystem is
unique. Solution of the Cauchy problem for the complete hydrodynamic system,
containing seven equations, is unique only for irrotational flows. For
vortical flows solution of the Cauchy problem is not unique. The reason of
the nonuniqueness is an interfusion, which cannot be taken into account
properly in the framework of hydrodynamics. There are some arguments in
favour of connection between interfusion and turbulence.
\end{abstract}

\section{Introduction}

Describing a fluid flow, one labels particles of the fluid by means of
Lagrangian coordinates $\mathbf{\xi =}\left\{ \xi _{1},\xi _{2},\xi
_{3}\right\} $. One supposes, that any Lagrangian coordinates $\mathbf{\xi }$
label the same fluid particle all the time. Applying laws of Newtonian
dynamics to any fluid particle, one obtains hydrodynamic equations for the
fluid flow in the Lagrangian representation (in the Lagrangian coordinates).
The Lagrangian representation is sensitive to the correct labeling of the
fluid particles in the sense, that there are situations, when in different
time moments the same Lagrangian coordinates $\mathbf{\xi }$ describe
different fluid particles. For instance, if two like gas beams, consisting
of noninteracting molecules, pass one through another, the particle labeling
changes after "collision" of the two beams. The picture is shown in the
figure, which describes world lines of gas particles in the space-time. The
solid lines shows gas particle with the same labeling $\mathbf{\xi }$,
whereas dashed lines show world lines of real gas molecules.

In this example a violation of the fluid particle labeling after "collision"
is evident. The stream lines, represented by solid lines, do not describe
motion of real fluid particles.

Is it important? Can we observe stream lines of a fluid? Let us imagine that
we introduce several flecks of dust in one of gas flows and follow their
motion. We suppose that the size of flecks is larger, than the size of gas
molecules, and any fleck interacts with many gas molecules. Then we may
think, that any fleck of dust moves along stream line of the fluid. As far
as collision of any fleck of dust with gas molecules is random, the flecks
of dust after "collision" appear in both gas beams, although before the
"collision" they were placed only in one of them. On the other hand, it is
generally assumed, that flecks of dust moves together with the gas, and any
fleck moves along the stream line of the fluid. Observation of flecks of
powder is a usual method of the stream lines investigation.

The hydrodynamic description of a fluid is valid at the supposition, that
one stream of a fluid can penetrate into the other one only to the depth of
the mean length of collision path. In this case the interfusion of different
streams of a fluid will be infinitesimal. However, such a small interfusion
will take place, and this interfusion may appear to be essential for the
shape of stream lines.

Such a physical phenomenon as turbulence can be discovered only, if one
traces the irregular behavior of stream lines. In other words, for
observation of turbulence a displacement of fluid particles is important,
but not only their velocities. The velocities are important only as a source
of displacement.

A motion of the inviscid barotropic fluid is described by the Euler equations

\begin{eqnarray}
\frac{\partial \mathbf{v}}{\partial t}+\left( \mathbf{v\nabla }\right) 
\mathbf{v} &=&-\frac{1}{\rho }\mathbf{\nabla }p,\qquad p=p\left( \rho
\right) =\rho ^{2}\frac{\partial E}{\partial \rho }  \label{a1.1} \\
\frac{\partial \rho }{\partial t}+\mathbf{\nabla }\left( \rho \mathbf{v}%
\right) &=&0  \label{a1.2}
\end{eqnarray}%
where $\rho =\rho \left( t,\mathbf{x}\right) $ is the fluid density, $%
\mathbf{v}=\mathbf{v}\left( t,\mathbf{x}\right) $ is the fluid velocity, $%
p=p\left( \rho \right) =\rho ^{2}\partial E\left( \rho \right) /\partial
\rho $ is the pressure and $E\left( \rho \right) $ is the fluid internal
energy per unit mass.

Stream lines are described by a system of ordinary differential equations.
One supposes that fluid particles move along the stream lines $\mathbf{x}=%
\mathbf{x}\left( t\right) $, which are defined by the equation

\begin{equation}
\frac{d\mathbf{x}}{dt}=\mathbf{v}\left( t,\mathbf{x}\right)  \label{a1.3}
\end{equation}%
where $\mathbf{v}=\mathbf{v}\left( t,\mathbf{x}\right) $ is a solution of
the Euler equations (\ref{a1.1}), (\ref{a1.2}).

The Euler system (\ref{a1.1}), (\ref{a1.2}) is a closed system of
differential equations, which may be solved independently of equations (\ref%
{a1.3}). The Euler system is a system of nonlinear partial differential
equations. It is difficult for solution. The system of ordinary differential
equations (\ref{a1.3}) is simpler, than the Euler system. Besides, it can be
solved only after solution of the Euler system (\ref{a1.1}), (\ref{a1.2}).
It is a reason, why researchers investigate mainly the Euler system. The
system of equations (\ref{a1.3}) for the stream lines remains usually
outside the region of consideration as some triviality.

However, the Euler system (\ref{a1.1}), (\ref{a1.2}) and equations (\ref%
{a1.3}) are dynamic equations of one dynamic system, and they should be
considered together. This dynamic system will be referred as the complete
hydrodynamic system. The Euler equations (\ref{a1.1}), (\ref{a1.2}) are not
dynamic equations of a wholesome dynamic system, because the cannot be
deduced from a variational principle, whereas dynamic equations (\ref{a1.1}%
), (\ref{a1.2}), (\ref{a1.3}) can. It seems, that dynamic equations do not
influence on the solution of dynamic equations (\ref{a1.1}), (\ref{a1.2}).
In reality, it is true only for irrotational flows. In vortical flows a
situation changes. In the vortical flows an interfusion appears. The
interfusion is not so large as in figure with colliding gas beams. This
interfusion is infinitesimal. It is conditioned by different velocities of
adjacent fluid volumes. This infinitesimal interfusion influences on the
shape of stream lines and on the labeling of the fluid particles, although
it does not influence on quantities $\rho ,\mathbf{v}$, which are solution
of the Euler system (\ref{a1.1}), (\ref{a1.2}).

In the case of the Euler representation of hydrodynamic equations, the fluid
particle labeling is not necessary. The four hydrodynamic Euler equations
are obtained as a result of the conservation laws of the energy and
momentum. However, in this case the conservation law of the angular momentum
is not used, and one cannot be sure, that the system of four Euler equations
for barotropic fluid describes completely rotational degrees of freedom of
molecules and those of fluid particles. The rotational degrees of freedom
may be essential in turbulent flows, where close consideration of rotational
degrees of freedom may appear to be essential.

In this paper we try to take into account interfusion, which appears in the
rotational flows of the barotropic fluid. Influence of interfusion manifests
itself in labeling of the fluid particles by means of the Lagrangian
coordinates $\mathbf{\xi }=\left\{ \xi _{1},\xi _{2},\xi _{3}\right\} $.
This change of labeling is not important for solution of Euler equations (%
\ref{a1.1}), (\ref{a1.2}), which form a closed system of differential
equations. However, this change of labeling may appear to be important for
such physical phenomena, where the shape of stream lines is essential (such
as turbulence).

We consider connection between the labeling and the interfusion on the
formal mathematical level. We shall consider the Euler dynamic equation for
inviscid barotropic fluid

We are interesting in the question, whether the labeling, generated by the
equation (\ref{a1.3}) is an unique possible way of labeling. To solve this
problem, we shall consider the Lagrangian coordinates $\mathbf{\xi }$ to be
dependent dynamic variables. The Eulerian coordinates $t,\mathbf{x}$ are
considered to be independent dynamic variables.

Thus, we consider dynamic system, described by seven dependent dynamic
variables $\rho ,\mathbf{v,\xi }$\textbf{, }which are functions of four
independent variables $t,\mathbf{x}$. Note that system of dynamic equations (%
\ref{a1.1}), (\ref{a1.2}) is a closed system of dynamic equations. However,
the dynamic system, described, by four dependent variables $\rho ,\mathbf{v}$
is not a wholesome dynamic system in the sense, that dynamic equations (\ref%
{a1.1}), (\ref{a1.2}) cannot be obtained from some a variational principle.

To obtain dynamic equations (\ref{a1.1}), (\ref{a1.2}) from the variational
principle \cite{S88}, one needs to add so-called Lin constraints \cite{L63}.
This conditions have the form 
\begin{equation}
\partial _{0}\mathbf{\xi +}\left( \mathbf{v\nabla }\right) \mathbf{\xi }=0
\label{b1.3}
\end{equation}%
It is easy to see, that characteristics of the linear differential equation (%
\ref{b1.3})%
\begin{equation}
\frac{dt}{1}=\frac{dx^{1}}{v^{1}}=\frac{dx^{2}}{v^{2}}=\frac{dx^{3}}{v^{3}}
\label{b1.5}
\end{equation}%
coincide with the equation (\ref{a1.3}). Vice versa, any integral of the
equation system (\ref{a1.3}) is a solution of the equation (\ref{b1.3}). The
Lin constraints (\ref{b1.3}) are interesting in the relation, that
independent dynamic variables in (\ref{b1.3}) are the same, as in the
dynamic equations (\ref{a1.1}), (\ref{a1.2}). Hence, dynamic equations (\ref%
{a1.1}), (\ref{a1.2}) and dynamic equations (\ref{b1.3}) may be considered
as dynamic equations of one dynamic system. It is rather difficult to
consider system of equations (\ref{a1.1}), (\ref{a1.2}), (\ref{a1.3}) as a
dynamic equations of a dynamic system, because independent variables are
different in equations (\ref{a1.1}), (\ref{a1.2}) and (\ref{a1.3}).

\section{Generalized stream function}

Let us note that the quantities $\mathbf{\xi }$ may be considered to be the
generalized stream function (GSF), because $\mathbf{\xi }$ have two main
properties of the stream function.

1. GSF $\mathbf{\xi }$ labels stream lines of a fluid.

2. Some combinations of the first derivatives of any $\mathbf{\xi }$ satisfy
the continuity equation identically.%
\begin{equation}
\partial _{k}j^{k}\equiv 0,\qquad j^{k}=\frac{\partial J_{\xi /x}}{\partial
\xi _{0,k}},\qquad \partial _{k}\equiv \frac{\partial }{\partial x^{k}}%
,\qquad k=0,1,2,3  \label{a2.1}
\end{equation}%
where $j^{k}=\left\{ j^{0},j^{1},j^{2},j^{3}\right\} =\left\{ \rho ,\rho 
\mathbf{v}\right\} $ is the 4-vector of flux. Here and in what follows, a
summation over two repeated indices is produced (0-3) for Latin indices and
(1-3) for Greek ones. The Jacobian determinant $J=J_{\xi /x}$ 
\begin{equation}
J_{\xi /x}=J\left( \xi _{l,k}\right) =\frac{\partial \left( \xi _{0}\mathbf{,%
}\xi _{1},\xi _{2},\xi _{3}\right) }{\partial \left(
x^{0},x^{1},x^{2},x^{3}\right) }=\det \left\vert \left\vert \xi
_{l,k}\right\vert \right\vert ,\qquad \xi _{l,k}\equiv \frac{\partial \xi
_{l}}{\partial x^{k}}\qquad l,k=0,1,2,3  \label{a2.2}
\end{equation}%
is considered to be a four-linear function of $\xi _{l,k}$. The quantity $%
\xi _{0}$ is the temporal Lagrangian coordinate, which appears to be
fictitious in expressions for the flux 4-vector $j^{k}$ 
\begin{eqnarray}
\rho &=&j^{0}=\frac{\partial \left( \xi _{1},\xi _{2},\xi _{3}\right) }{%
\partial \left( x^{1},x^{2},x^{3}\right) },\qquad \rho v^{1}=j^{1}=-\frac{%
\partial \left( \xi _{1},\xi _{2},\xi _{3}\right) }{\partial \left(
t,x^{2},x^{3}\right) },  \nonumber \\
\rho v^{2} &=&j^{2}=\frac{\partial \left( \xi _{1},\xi _{2},\xi _{3}\right) 
}{\partial \left( t,x^{1},x^{3}\right) },\qquad \rho v^{3}=j^{3}=-\frac{%
\partial \left( \xi _{1},\xi _{2},\xi _{3}\right) }{\partial \left(
t,x^{1},x^{2}\right) }  \label{a2.3}
\end{eqnarray}

A use of Jacobians in the description of the ideal fluid goes up to Clebsch 
\cite[Clebsch, 1857,1859]{C57,C59}, who used Jacobians in the expanded form.
It was rather bulky. We use a more rational designations, when the 4-flux
and other essential dynamic quantities are presented in the form of
derivatives of the principal Jacobian $J$. Dealing with the generalized
stream function $\xi =\left\{ \xi _{0},\xi _{1},\xi _{2},\xi _{3}\right\} $,
the following identities are useful 
\begin{equation}
\frac{\partial J}{\partial \xi _{i,l}}\xi _{k,l}\equiv J\delta
_{k}^{i},\qquad \partial _{k}\frac{\partial J}{\partial \xi _{0,k}}\equiv
0,\qquad \partial _{l}\frac{\partial ^{2}J}{\partial \xi _{0,k}\partial \xi
_{i,l}}\equiv 0  \label{a2.4}
\end{equation}%
\begin{equation}
\frac{\partial ^{2}J}{\partial \xi _{0,k}\partial \xi _{l,s}}\equiv
J^{-1}\left( \frac{\partial J}{\partial \xi _{0,k}}\frac{\partial J}{%
\partial \xi _{l,s}}-\frac{\partial J}{\partial \xi _{0,s}}\frac{\partial J}{%
\partial \xi _{l,k}}\right)  \label{a2.5}
\end{equation}%
See details of working with Jacobians and the generalized stream functions
in \cite[Rylov,2004]{R2004}.

\textit{Example}.\textit{\ Application of the stream function for
integration of equations, describing the 2D stationary flow of
incompressible fluid.}

Dynamic equations have the form%
\begin{equation}
u_{x}+v_{y}=0,\qquad \partial _{y}\left( uu_{x}+vu_{y}\right) =\partial
_{x}\left( uv_{x}+vv_{y}\right)  \label{a2.6}
\end{equation}%
where $u$ and $v$ are velocity components along $x$-axis and $y$-axis
respectively.

Introducing the stream function $\psi $ by means of relations%
\begin{equation}
u=-\psi _{y},\qquad v=\psi _{x}  \label{a2.7}
\end{equation}%
we satisfy the first equation (\ref{a2.6}) identically, and we obtain for
the second equation (\ref{a2.6}) the relations%
\begin{eqnarray*}
\psi _{y}\psi _{xyy}-\psi _{x}\psi _{yyy} &=&-\psi _{y}\psi _{xxx}+\psi
_{x}\psi _{xxy} \\
\psi _{y}\left( \psi _{xyy}+\psi _{xxx}\right) &=&\psi _{x}\left( \psi
_{xxy}+\psi _{yyy}\right)
\end{eqnarray*}%
which can be rewritten in the form%
\begin{equation}
\frac{\partial \left( \omega ,\psi \right) }{\partial \left( x,y\right) }%
=0,\qquad \omega \equiv \psi _{xx}+\psi _{yy}  \label{a2.8}
\end{equation}%
where $\omega $ is the vorticity of the fluid flow. The general solution of
equation (\ref{a2.8}) has the form%
\begin{equation}
\omega =\psi _{xx}+\psi _{yy}=\Omega \left( \psi \right)  \label{a2.9}
\end{equation}%
where $\Omega $ is an arbitrary function of $\psi $.

For the irrotational flow the vorticity $\Omega \left( \psi \right) =0$, and
we obtain instead (\ref{a2.9}) 
\begin{equation}
\psi _{xx}+\psi _{yy}=0  \label{a2.10}
\end{equation}

One obtains the unique solution of (\ref{a2.10}) inside of a closed region
of 2D space provided, that the value $\psi |_{\Sigma }$ of the stream
function $\psi $ is given on the boundary $\Sigma $ of this region. The
differential structure of equations (\ref{a2.9}) and (\ref{a2.10}) is
similar. One should expect, that giving the value $\psi |_{\Sigma }$ of the
stream function $\psi $ on the boundary $\Sigma $, one obtains the unique
solution of the equation (\ref{a2.10}). But it is not so, because the
indefinite function $\Omega \left( \psi \right) $ is not given, and it
cannot be determined from the boundary condition, because the nature of the
function $\Omega \left( \psi \right) $ is another, than the nature of the
boundary conditions. First, if the flow contains closed stream lines, which
do not cross the boundary, one cannot determine the values of $\Omega $ on
these stream lines from the boundary conditions. But for determination of
the unique solution the values of $\Omega $ on the closed stream lines must
be given. Second, the boundary conditions are given arbitrarily. The
function $\Omega $ cannot be given arbitrarily. For those stream lines,
which cross the boundary more than once, the values of $\Omega $ on the
different segments of the boundary are to be agreed. Thus, the nonuniqueness
of the solution, connected with the indefinite function $\Omega $ has
another nature, than the nonuniqueness, connected with the insufficiency of
the boundary conditions.

\section{Derivation of hydrodynamic equations from the variational principle}

We use the variational principle for the derivation of the hydrodynamic
equations (\ref{a1.1}), (\ref{a1.2}), (\ref{b1.3}). The action functional
has the form%
\begin{equation}
\mathcal{A}\left[ \xi ,j,p\right] =\dint\limits_{V_{x}}\left\{ \frac{\mathbf{%
j}^{2}}{2\rho }-\rho E\left( \rho \right) -p_{k}\left( j^{k}-\rho _{0}\left( 
\mathbf{\xi }\right) \frac{\partial J}{\partial \xi _{0,k}}\right) \right\}
d^{4}x,  \label{a3.1}
\end{equation}%
where $p_{k}$, $k=0,1,2,3$ are the Lagrange multipliers, which introduce the
designations for the 4-flux%
\begin{equation}
j^{k}=\rho _{0}\left( \mathbf{\xi }\right) \frac{\partial J}{\partial \xi
_{0,k}},\qquad k=0,1,2,3  \label{a3.2}
\end{equation}%
Note, the expression for the 4-flux (\ref{a3.2}) satisfies the first
equation (\ref{a2.1}) identically, because the expression (\ref{a3.2}) may
be reduced to the form of the second relation (\ref{a2.1}) by means of a
change of variables $\mathbf{\xi }$%
\[
\tilde{\xi}_{0}=\xi _{0},\qquad \tilde{\xi}_{1}=\int \rho _{0}\left( \mathbf{%
\xi }\right) d\xi _{1},\qquad \tilde{\xi}_{2}=\xi _{2},\qquad \tilde{\xi}%
_{3}=\xi _{3} 
\]%
Then%
\[
\rho _{0}\left( \mathbf{\xi }\right) \frac{\partial \left( \xi _{0}\mathbf{,}%
\xi _{1},\xi _{2},\xi _{3}\right) }{\partial \left(
x^{0},x^{1},x^{2},x^{3}\right) }=\frac{\partial \left( \tilde{\xi}_{0}%
\mathbf{,}\tilde{\xi}_{1},\tilde{\xi}_{2},\tilde{\xi}_{3}\right) }{\partial
\left( x^{0},x^{1},x^{2},x^{3}\right) },\qquad \tilde{\xi}_{1}=\int \rho
_{0}\left( \mathbf{\xi }\right) d\xi _{1} 
\]

Besides according to the first identity (\ref{a2.4}) the relation (\ref{a3.2}%
) satisfies the Lin constraint (\ref{b1.3}).

Variation of the action (\ref{a3.1}) with respect to $p_{k}=\left\{ p_{0},%
\mathbf{p}\right\} $ gives relations (\ref{a3.2}). Another dynamic equations
have the form%
\begin{equation}
\delta \rho :\qquad p_{0}=-\frac{\mathbf{j}^{2}}{2\rho ^{2}}-\frac{\partial 
}{\partial \rho }\left( \rho E\left( \rho \right) \right) =-\frac{\mathbf{v}%
^{2}}{2}-\frac{\partial }{\partial \rho }\left( \rho E\right)  \label{a3.3}
\end{equation}%
\begin{equation}
\delta \mathbf{j:}\qquad \mathbf{p}=\frac{\mathbf{j}}{\rho }=\mathbf{v}
\label{a3.4}
\end{equation}%
\begin{equation}
\delta \xi _{l}:\qquad -\partial _{s}\left( \rho _{0}\left( \mathbf{\xi }%
\right) p_{k}\frac{\partial ^{2}J}{\partial \xi _{0,k}\partial \xi _{l,s}}%
\right) +p_{k}\frac{\partial \rho _{0}}{\partial \xi _{l}}\left( \mathbf{\xi 
}\right) \frac{\partial J}{\partial \xi _{0,k}}=0,\qquad l=0,1,2,3
\label{a3.5}
\end{equation}%
Using the third relation (\ref{a2.4}), we obtain%
\begin{equation}
-\frac{\partial ^{2}J}{\partial \xi _{0,k}\partial \xi _{l,s}}\left( \frac{%
\partial \rho _{0}\left( \mathbf{\xi }\right) }{\partial \xi _{\alpha }}\xi
_{\alpha ,s}+\rho _{0}\left( \mathbf{\xi }\right) \partial _{s}p_{k}\right)
+p_{k}\frac{\partial \rho _{0}}{\partial \xi _{l}}\left( \mathbf{\xi }%
\right) \frac{\partial J}{\partial \xi _{0,k}}=0  \label{a3.6}
\end{equation}

Now using (\ref{a2.5}), we obtain 
\begin{equation}
-J^{-1}\left( \frac{\partial J}{\partial \xi _{0,k}}\frac{\partial J}{%
\partial \xi _{l,s}}-\frac{\partial J}{\partial \xi _{0,s}}\frac{\partial J}{%
\partial \xi _{l,k}}\right) \left( \frac{\partial \rho _{0}\left( \mathbf{%
\xi }\right) }{\partial \xi _{\alpha }}\xi _{\alpha ,s}+\rho _{0}\left( 
\mathbf{\xi }\right) \partial _{s}p_{k}\right) +p_{k}\frac{\partial \rho _{0}%
}{\partial \xi _{l}}\left( \mathbf{\xi }\right) \frac{\partial J}{\partial
\xi _{0,k}}=0  \label{a3.7}
\end{equation}%
Using the first relation (\ref{a2.4}), we obtain%
\begin{equation}
J^{-1}\left( \frac{\partial J}{\partial \xi _{0,k}}\frac{\partial J}{%
\partial \xi _{l,s}}-\frac{\partial J}{\partial \xi _{0,s}}\frac{\partial J}{%
\partial \xi _{l,k}}\right) \rho _{0}\left( \mathbf{\xi }\right) \partial
_{s}p_{k}=0,\qquad l=0,1,2,3  \label{a3.8}
\end{equation}

There are two ways of dealing with this equation:

\noindent 1. Elimination of GSF $\mathbf{\xi }$, which leads to the Euler
equations.

\noindent 2. Integration, which leads to appearance of arbitrary functions.

\textit{The first way: elimination of GSF}

Convoluting (\ref{a3.8}) with $\xi _{l,i}$ and using dynamic equations (\ref%
{a3.2}), we obtain 
\begin{equation}
j^{k}\partial _{i}p_{k}-j^{k}\partial _{k}p_{i}=0,\qquad i=0,1,2,3
\label{a3.9}
\end{equation}

Substituting $p_{k}$ and $j^{k}$ from relations (\ref{a3.3}) and (\ref{a3.4}%
), we obtain the Euler dynamic equations (\ref{a1.1})%
\begin{equation}
\partial _{0}v^{\alpha }+\left( \mathbf{v\nabla }\right) v^{\alpha
}=-\partial _{\alpha }\frac{\partial }{\partial \rho }\left( \rho E\right) =-%
\frac{1}{\rho }\partial _{\alpha }p,\qquad \alpha =1,2,3,\qquad p=\rho ^{2}%
\frac{\partial E}{\partial \rho }  \label{a3.10}
\end{equation}%
The continuity equation (\ref{a1.2}) is a corollary of equations (\ref{a3.2}%
) and identity (\ref{a2.1}). Finally the Lin constraints (\ref{b1.3}) are
corollaries of the first identity (\ref{a2.4}) and dynamic equations (\ref%
{a3.2}).

\textit{The second way: integration of the equation for } $p_{k}$

Let us consider the equations (\ref{a3.8}) as linear differential equations
for $p_{k}$. The general solution of (\ref{a3.8}) has the form%
\begin{equation}
p_{k}=\left( \partial _{k}\varphi +g^{\alpha }\left( \mathbf{\xi }\right)
\partial _{k}\xi _{\alpha }\right) ,\qquad k=0,1,2,3  \label{a3.11}
\end{equation}%
where $g^{\alpha }\left( \mathbf{\xi }\right) ,$ $\alpha =1,2,3$ are
arbitrary functions of $\mathbf{\xi }$, $\varphi =g^{0}\left( \xi
_{0}\right) $ is a new variable instead of fictitious variable $\xi _{0}$.
Let us differentiate (\ref{a3.11}) and substitute the obtained expressions 
\begin{equation}
\partial _{s}p_{k}=\left( \partial _{s}\partial _{k}\varphi +g^{\alpha
}\left( \mathbf{\xi }\right) \partial _{s}\partial _{k}\xi _{\alpha }\right)
+\frac{\partial g^{\alpha }\left( \mathbf{\xi }\right) }{\partial \xi
_{\beta }}\partial _{k}\xi _{\alpha }\partial _{s}\xi _{\beta }
\label{a3.12}
\end{equation}%
in (\ref{a3.8}). Using the first identity (\ref{a2.4}), we see, that the
relations (\ref{a3.12}) satisfy the equations (\ref{a3.8}) identically.

We may substitute (\ref{a3.11}) in the action (\ref{a3.1}), or introduce (%
\ref{a3.11}) by means of the Lagrange multipliers. (the result is the same).
We obtain the new action functional 
\begin{equation}
\mathcal{A}\left[ \xi ,j\right] =\dint\limits_{V_{x}}\left\{ \frac{\mathbf{j}%
^{2}}{2\rho }-\rho E\left( \rho \right) -j^{k}\left( \partial _{k}\varphi
+g^{\alpha }\left( \mathbf{\xi }\right) \partial _{k}\xi _{\alpha }\right)
\right\} d^{4}x,  \label{a3.14}
\end{equation}%
which contains arbitrary integration functions $g\left( \mathbf{\xi }\right) 
$. Here%
\begin{equation}
j^{0}=\rho ,\qquad \mathbf{j}=\rho \mathbf{v=}\left\{
j^{1},j^{2},j^{3}\right\}  \label{a3.15}
\end{equation}%
The integration functions $g\left( \mathbf{\xi }\right) $ are considered as
a fixed functions of $\mathbf{\xi }$. The term 
\begin{equation}
p_{k}\frac{\partial J}{\partial \xi _{0,k}}=\left( \partial _{k}\varphi
+g^{\alpha }\left( \mathbf{\xi }\right) \partial _{k}\xi _{\alpha }\right) 
\frac{\partial J}{\partial \xi _{0,k}}=\frac{\partial \left( \varphi \mathbf{%
,}\xi _{1},\xi _{2},\xi _{3}\right) }{\partial \left(
x^{0},x^{1},x^{2},x^{3}\right) }  \label{a3.16}
\end{equation}%
is omitted, because it does not contribute to dynamic equations.

Variation of (\ref{a3.14}) with respect to $\varphi $, $\rho $ and $j^{\mu }$
gives respectively%
\begin{equation}
\delta \varphi :\qquad \partial _{k}j^{k}=0  \label{a3.17}
\end{equation}%
\begin{equation}
\delta \rho :\qquad \partial _{0}\varphi +g^{\beta }\left( \mathbf{\xi }%
\right) \partial _{0}\xi _{\beta }+\frac{\mathbf{j}^{2}}{2\rho ^{2}}+\frac{%
\partial }{\partial \rho }\left( \rho E\left( \rho \right) \right) =0
\label{a3.18}
\end{equation}%
\begin{equation}
\delta j^{\mu }:\qquad v^{\mu }\equiv \frac{j^{\mu }}{\rho }=\partial _{\mu
}\varphi +g^{\alpha }\left( \mathbf{\xi }\right) \partial _{\mu }\xi
_{\alpha }  \label{a3.19}
\end{equation}

Variation of (\ref{a3.14}) with respect to $\xi _{\alpha }$ gives%
\begin{equation}
\delta \xi _{\alpha }:\qquad \rho \Omega ^{a\mu }\left( \mathbf{\xi }\right)
\left( \partial _{0}\xi _{\alpha }+\left( \mathbf{v\nabla }\right) \xi
_{\alpha }\right) =0,  \label{a3.20}
\end{equation}%
where%
\begin{equation}
\Omega ^{a\mu }\left( \mathbf{\xi }\right) =\left( \frac{\partial g^{\alpha
}\left( \mathbf{\xi }\right) }{\partial \xi _{\mu }}-\frac{\partial g^{\mu
}\left( \mathbf{\xi }\right) }{\partial \xi _{\alpha }}\right)  \label{a3.21}
\end{equation}%
and $\mathbf{v}$ is determined by the relation (\ref{a3.19})

If $\det \left\vert \left\vert \Omega ^{\alpha \beta }\right\vert
\right\vert \neq 0$, then the Lin constraints 
\begin{equation}
\left( \partial _{0}\xi _{\alpha }+\left( \mathbf{v\nabla }\right) \xi
_{\alpha }\right) =0  \label{a3.22}
\end{equation}%
follows from (\ref{a3.20})

However, the matrix $\Omega ^{\alpha \beta }$ is antisymmetric and 
\begin{equation}
\det \left\vert \left\vert \Omega ^{\alpha \beta }\right\vert \right\vert
=\left\vert 
\begin{array}{ccc}
0 & \Omega ^{12} & \Omega ^{13} \\ 
-\Omega ^{12} & 0 & \Omega ^{23} \\ 
-\Omega ^{13} & -\Omega ^{23} & 0%
\end{array}%
\right\vert \equiv 0  \label{a3.23}
\end{equation}%
Then it follows from (\ref{a3.20})%
\begin{equation}
\partial _{0}\xi _{\alpha }+\left( \mathbf{v\nabla }\right) \xi _{\alpha }=-%
\frac{\omega }{\rho _{0}\left( \mathbf{\xi }\right) }\varepsilon _{\alpha
\beta \gamma }\Omega ^{\beta \gamma }\left( \mathbf{\xi }\right) \qquad
\alpha =1,2,3  \label{a3.24}
\end{equation}%
where $\omega =\omega \left( t,\mathbf{\xi }\right) $ is an arbitrary
quantity, and $\rho _{0}\left( \mathbf{\xi }\right) $ is the weight function
from (\ref{a3.2}).

The obtained equation (\ref{a3.24}) contains the initial dynamic equation (%
\ref{b1.3}) as a special case. For irrotational flow, when $\Omega ^{\beta
\gamma }\left( \mathbf{\xi }\right) =0$, the equation (\ref{a3.24}) turns to
(\ref{b1.3}). In the action functional (\ref{a3.1}) the initial relation (%
\ref{b1.3}) is used as a side constraint. It is a reason, why the equation (%
\ref{a3.24}) is not obtained from the action functional (\ref{a3.1}).

Note, that eliminating the variables $\varphi $ and $\mathbf{\xi }$ from
dynamic equations (\ref{a3.18}) - (\ref{a3.20}), we obtain the Euler dynamic
equations (\ref{a1.1}).

The vorticity $\mathbf{\omega }_{0}\equiv \nabla \times \mathbf{v}$ and $%
\mathbf{v}\times \mathbf{\omega }_{0}$ are obtained from (\ref{a3.19}) in
the form 
\begin{equation}
\mathbf{\omega }_{0}=\mathbf{\nabla }\times \mathbf{v}={\frac{1}{2}}{\Omega }%
^{\alpha \beta }\mathbf{\nabla }\xi _{\beta }\times \mathbf{\nabla }\xi
_{\alpha }  \label{b3.28}
\end{equation}%
\begin{equation}
\mathbf{v}\times \mathbf{\omega }_{0}={\Omega }^{\alpha \beta }\mathbf{%
\nabla }\xi _{\beta }(\mathbf{v\nabla })\xi _{\alpha }  \label{b3.29}
\end{equation}

Let us form a difference between the time derivative of (\ref{a3.19}) and
the gradient of (\ref{a3.18}). Eliminating $\Omega ^{a\mu }\left( \mathbf{%
\xi }\right) \partial _{0}\mathbf{\xi }_{\alpha }$ from the obtained
equation by means of equations (\ref{a3.20}), one obtains 
\begin{equation}
\partial _{0}\mathbf{v}+\mathbf{\nabla }\frac{\mathbf{v}^{2}}{2}+\frac{%
\partial ^{2}(\rho E)}{\partial \rho ^{2}}\mathbf{\nabla }\rho -\Omega
^{\alpha \beta }\mathbf{\nabla }\xi _{\beta }(\mathbf{v\nabla })\xi _{\alpha
}=0  \label{b3.30}
\end{equation}%
Using (\ref{b3.29}) and (\ref{b3.28}), the expression (\ref{b3.30}) reduces
to 
\begin{equation}
\partial _{0}\mathbf{v}+\mathbf{\nabla }{\frac{\mathbf{v}^{2}}{2}}+\frac{1}{%
\rho }\mathbf{\nabla }(\rho ^{2}\frac{\partial E}{\partial \rho })-\mathbf{v}%
\times (\mathbf{\nabla }\times \mathbf{v})=0  \label{b3.31}
\end{equation}%
In virtue of the identity 
\begin{equation}
\mathbf{v}\times (\mathbf{\nabla }\times \mathbf{v})\equiv \mathbf{\nabla }{%
\frac{\mathbf{v}^{2}}{2}}-(\mathbf{v\nabla })\mathbf{v}  \label{b3.32}
\end{equation}%
the equation (\ref{b3.31}) is equivalent to (\ref{a1.1}).

Note, that the Euler equations (\ref{a1.1}) are obtained at any form of the
arbitrary function $\omega \left( t,\mathbf{\xi }\right) $ in the equations (%
\ref{a3.24}), because the equations (\ref{a3.24}) are used in the form (\ref%
{a3.20}), where the form of $\omega \left( t,\mathbf{\xi }\right) $ is
unessential. Solution of the Euler system (\ref{a1.1}), (\ref{a1.2}) in the
form $\rho =\rho \left( t,\mathbf{x}\right) $, $\mathbf{v}=\mathbf{v}\left(
t,\mathbf{x}\right) $ does not depend on the form of the indefinite function 
$\omega \left( t,\mathbf{\xi }\right) $.

If $\omega \left( t,\mathbf{\xi }\right) \neq 0$, the dynamic equations (\ref%
{a3.24}) describe a violation of the Lin constraints (\ref{b1.3}). One
obtains another labeling of the stream lines, than that one, which is
described by the Lin constraints (\ref{b1.3}). If the flow is irrotational,
and $\mathbf{\Omega }=0$, the labeling does not depend on $\omega \left( t,%
\mathbf{\xi }\right) $.

Let us consider two different labeling $\mathbf{\xi }$ and $\mathbf{\bar{\xi}%
}$ of the the same fluid flow described by the variables $\rho =\rho \left(
t,\mathbf{x}\right) $, $\mathbf{v}=\mathbf{v}\left( t,\mathbf{x}\right) $.
The initial conditions are supposed to have the form%
\begin{eqnarray}
\varphi \left( 0,\mathbf{x}\right) &=&\varphi _{\mathrm{in}}\left( \mathbf{x}%
\right) =0,\qquad \rho \left( 0,\mathbf{x}\right) =\rho _{\mathrm{in}}\left( 
\mathbf{x}\right) ,  \label{a3.30} \\
\bar{\xi}_{\alpha }\left( 0,\mathbf{x}\right) &=&\left( \bar{\xi}_{\alpha
}\right) _{\mathrm{in}}\left( \mathbf{x}\right) =x^{\alpha },\qquad \alpha
=1,2,3  \label{a3.30c}
\end{eqnarray}%
Then according to (\ref{a3.19}), (\ref{a3.21})%
\begin{equation}
g^{\alpha }\left( \mathbf{\bar{\xi}}_{\mathrm{in}}\left( \mathbf{x}\right)
\right) =g^{\alpha }\left( \mathbf{x}\right) =v_{\mathrm{in}}^{\alpha
}\left( \mathbf{x}\right) ,\qquad \alpha =1,2,3  \label{a3.30a}
\end{equation}%
According to (\ref{a3.24}) the dynamic equations for labeling $\mathbf{\xi }$
and $\mathbf{\bar{\xi}}$ have the form 
\begin{equation}
\partial _{0}\xi _{\alpha }+\left( \mathbf{v\nabla }\right) \xi _{\alpha
}=0\qquad \alpha =1,2,3  \label{b3.40}
\end{equation}%
\begin{equation}
\partial _{0}\bar{\xi}_{\alpha }+\left( \mathbf{v\nabla }\right) \bar{\xi}%
_{\alpha }=-\frac{\omega }{\rho _{0}\left( \mathbf{\bar{\xi}}\right) }%
\varepsilon _{\alpha \beta \gamma }\Omega ^{\beta \gamma }\left( \mathbf{%
\bar{\xi}}\right) \qquad \alpha =1,2,3  \label{b3.41}
\end{equation}%
where the velocity $\mathbf{v=v}\left( t,\mathbf{x}\right) $ is the same in
both equations and the function $\Omega ^{\beta \gamma }\left( \mathbf{x}%
\right) $ is defined by the relation 
\begin{equation}
\Omega ^{a\mu }\left( \mathbf{x}\right) =\Omega ^{a\mu }\left( \mathbf{\bar{%
\xi}}_{\mathrm{in}}\left( \mathbf{x}\right) \right) =\left( \frac{\partial
g^{\alpha }\left( \mathbf{\bar{\xi}}_{\mathrm{in}}\right) }{\partial \left( 
\bar{\xi}_{\mathrm{in}}\right) _{\mu }}-\frac{\partial g^{\mu }\left( 
\mathbf{\bar{\xi}}_{\mathrm{in}}\right) }{\partial \left( \bar{\xi}_{\mathrm{%
in}}\right) _{\alpha }}\right) =\left( \frac{\partial v_{\mathrm{in}%
}^{\alpha }\left( \mathbf{x}\right) }{\partial x^{\mu }}-\frac{\partial v_{%
\mathrm{in}}^{\mu }\left( \mathbf{x}\right) }{\partial x^{\alpha }}\right)
\label{a3.30b}
\end{equation}%
If the velocity $\mathbf{v}$ is defined by relations (\ref{a3.19}), it
satisfies the Euler equations and associates with the generalized stream
function $\mathbf{\xi }\left( t,\mathbf{x}\right) $, whose evolution is
described by the equations (\ref{a3.24})

In general, the evolution of the quantities $\mathbf{\bar{\xi}}$ and $%
\mathbf{\xi }$ is different, although the coincide at $t=0$. Let 
\begin{equation}
\mathbf{\eta }=\mathbf{\bar{\xi}-\xi }  \label{a3.32}
\end{equation}%
It follows from (\ref{b3.40}) and (\ref{b3.41}) that mismatch $\mathbf{\eta }
$ between $\mathbf{\bar{\xi}}$ and $\mathbf{\xi }$ is determined by the
relation%
\begin{eqnarray}
\partial _{0}\eta _{\alpha }+\mathbf{v}\left( t,\mathbf{x}\right) \mathbf{%
\nabla }\eta _{\alpha }+\frac{\omega \left( t,\mathbf{\bar{\xi}}\left( t,%
\mathbf{x}\right) \right) }{\rho _{0}\left( \mathbf{\bar{\xi}}\left( t,%
\mathbf{x}\right) \right) }\varepsilon _{\alpha \beta \gamma }\Omega ^{\beta
\gamma }\left( \mathbf{\bar{\xi}}\left( t,\mathbf{x}\right) \right)
&=&0,\qquad \eta _{\alpha }\left( 0,\mathbf{x}\right) =0,  \label{a3.33} \\
\alpha &=&1,2,3  \nonumber
\end{eqnarray}%
The system of ordinary differential equations, associated with the equation (%
\ref{a3.33}), has the form%
\begin{eqnarray}
\frac{dt}{1} &=&\frac{dx^{1}}{v^{1}\left( t,\mathbf{x}\right) }=\frac{dx^{2}%
}{v^{2}\left( t,\mathbf{x}\right) }=\frac{dx^{3}}{v^{3}\left( t,\mathbf{x}%
\right) }=\frac{\rho _{0}\left( \mathbf{\bar{\xi}}\left( t,\mathbf{x}\right)
\right) d\eta _{\alpha }}{\omega \left( t,\mathbf{\bar{\xi}}\left( t,\mathbf{%
x}\right) \right) \varepsilon _{\alpha \beta \gamma }\Omega ^{\beta \gamma
}\left( \mathbf{\bar{\xi}}\left( t,\mathbf{x}\right) \right) },
\label{a3.34} \\
\alpha &=&1,2,3  \nonumber
\end{eqnarray}

Solution of the system of ordinary equations at the initial conditions $%
\mathbf{\eta }\left( 0,\mathbf{x}\right) =0$ has the form%
\begin{equation}
\eta _{\mu }\left( t,\mathbf{x}\right) =\dint\limits_{0}^{t}\frac{\omega
\left( t,\mathbf{\bar{\xi}}\left( t,\mathbf{x}\right) \right) \varepsilon
_{\mu \beta \gamma }}{\rho _{0}\left( \mathbf{\bar{\xi}}\left( t,\mathbf{x}%
\right) \right) }\Omega ^{\beta \gamma }\left( \mathbf{\bar{\xi}}\left( t,%
\mathbf{x}\right) \right) dt,\qquad \mu =1,2,3  \label{a3.35}
\end{equation}%
Then%
\begin{equation}
\xi _{\mu }\left( t,\mathbf{x}\right) =\bar{\xi}_{\mu }\left( t,\mathbf{x}%
\right) -\dint\limits_{0}^{t}\frac{\omega \left( t,\mathbf{\bar{\xi}}\left(
t,\mathbf{x}\right) \right) \varepsilon _{\mu \beta \gamma }}{\rho
_{0}\left( \mathbf{\bar{\xi}}\left( t,\mathbf{x}\right) \right) }\Omega
^{\beta \gamma }\left( \mathbf{\bar{\xi}}\left( t,\mathbf{x}\right) \right)
dt,\qquad \mu =1,2,3  \label{a3.36}
\end{equation}

Thus, although solution $\rho ,\mathbf{v}$ of the Cauchy problem for the
Euler system of hydrodynamic equation (\ref{a1.1}), (\ref{a1.2}) is unique,
the solution $\rho ,\mathbf{v,\xi }$ for the Cauchy problem of the complete
system of hydrodynamic equations (\ref{a1.1}), (\ref{a1.2}), (\ref{a3.24})
is not unique. The reason of this nonuniqueness is consideration of
interfusion. This consideration is formal. One cannot understand mechanism
of the interfusion influence from this consideration. Nevertheless this
influence takes place, and it should be investigated more closely.

\section{Two-dimensional vortical flow of ideal barotropic fluid in the
three-dimensional space.}

It seems, that in the two-dimensional flow instead of determinant (\ref%
{a3.23}) we have the determinant%
\begin{equation}
\left\vert \left\vert 
\begin{array}{ll}
0 & \Omega ^{12} \\ 
-\Omega ^{12} & 0%
\end{array}%
\right\vert \right\vert =\left( \Omega ^{12}\right) ^{2}  \label{a4.1}
\end{equation}%
which does not vanish, in general. Then the problem of nonuniqueness of
thelabeling is removed and the solution of the Cauchy problem for the
complete hydrodynamic system becomes to be unique.

In reality, we may control the solution only via initial conditions. We may
give the two-dimensional initial conditions, i.e. 
\begin{equation}
\partial _{3}\mathbf{v}_{\mathrm{in}}\left( \mathbf{x}\right) =0,\qquad
\partial _{3}\rho _{\mathrm{in}}\left( \mathbf{x}\right) =0,\qquad v_{%
\mathrm{in}}^{3}\left( \mathbf{x}\right) =0  \label{a4.2}
\end{equation}%
In this case%
\begin{equation}
\Omega _{\mathrm{in}}^{12}\left( \xi _{1},\xi _{2}\right) =\frac{\partial v_{%
\mathrm{in}}^{1}\left( \xi _{1},\xi _{2}\right) }{\partial \xi _{2}}-\frac{%
\partial v_{\mathrm{in}}^{2}\left( \xi _{1},\xi _{2}\right) }{\partial \xi
_{1}},\qquad \Omega _{\mathrm{in}}^{23}\left( \mathbf{\xi }\right) =0,\qquad
\Omega _{\mathrm{in}}^{31}\left( \mathbf{\xi }\right) =0  \label{a4.3}
\end{equation}%
The determinant%
\begin{equation}
\det \left\vert \left\vert \Omega ^{\alpha \beta }\right\vert \right\vert
=\left\vert 
\begin{array}{ccc}
0 & \Omega ^{12} & 0 \\ 
-\Omega ^{12} & 0 & 0 \\ 
0 & 0 & 0%
\end{array}%
\right\vert \equiv 0  \label{a4.4}
\end{equation}%
and the relations (\ref{a3.24}) take the form%
\begin{eqnarray}
\partial _{0}\xi _{1}+\mathbf{v}\left( t,\mathbf{x}\right) \mathbf{\nabla }%
\xi _{1} &=&0,\qquad \partial _{0}\xi _{2}+\mathbf{v}\left( t,\mathbf{x}%
\right) \mathbf{\nabla }\xi _{2}=0  \label{a4.5} \\
\partial _{0}\xi _{3}+\mathbf{v}\left( t,\mathbf{x}\right) \mathbf{\nabla }%
\xi _{3} &=&-\frac{\omega \left( t,\mathbf{\xi }\right) }{\rho _{0}\left(
\xi _{1},\xi _{2}\right) }\Omega ^{1,2}\left( \xi _{1},\xi _{2}\right)
\label{a4.6}
\end{eqnarray}%
One cannot control indefinite quantity $\omega \left( t,\mathbf{\xi }\right) 
$, which may depend on $x^{3}$. The equation (\ref{a4.6}) generates
nonunique solution of the Cauchy problem of vortical flow for the complete
hydrodynamic system. The flow with the two-dimensional initial conditions
turns into three-dimensional vortical flow.

\section{Concluding remarks}

Solution of the Cauchy problem for the vortical flow of inviscid barotropic
fluid is not unique, if we solve seven dynamic equations of the complete
hydrodynamic system, which includes description of the shape of stream lines
and has seven dependent variables $\rho $, $\mathbf{v}$, $\mathbf{\xi }$.
Nonuniqueness is connected with the fact, the initial conditions for
variables $\rho $, $\mathbf{v}$, $\mathbf{\xi }$ do not control the
intermixing effect.

Solution of the Cauchy problem for the vortical flow of inviscid barotropic
fluid is unique, if we solve only four dynamic equations of the Euler system
and ignore shape of stream lines. In this case dynamic equations are written
for four dependent variables $\rho $, $\mathbf{v}$.

Solution of the Cauchy problem for the irrotational flow of inviscid
barotropic fluid is unique for both seven equations of the complete
hydrodynamic system and for four equations of the Euler system.

The intermixing effect, generating nonunique solutions, associates with the
turbulence phenomenon at the following points: (1) both effects are not
controlled by the initial data for variables $\rho $, $\mathbf{v}$, $\mathbf{%
\xi }$ of the hydrodynamic equations in the Euler representation, (2) both
effects take place at the vortical flows, and they are absent at the
irrotational flows, (3) both effects are strong at vanishing viscosity. We
admit, that the interfusion may be connected with the turbulence phenomena,
although we do not yet insist on this statement.

If, indeed, the turbulent phenomena are connected with the interfusion and
with the shape of stream lines, it becomes clear, why numerous
investigations of hydrodynamic equations, describing only density and
velocity, but not the shape of stream lines, had not led to a progress. The
researchers looked for turbulence in that region, where it is not placed.

CAPTIONS

Figure 1. Dashed lines show real trajectories of particles. The solid lines
show trajectories of the mean particle motion.


\begin{thebibliography}{9}
\bibitem{C57} A. Clebsch, \"{U}ber eine allgemaine Transformation der
hydrodynamischen Gleichungen, \textit{J. reine angew. Math.} \textbf{54 },
293-312 (1857).

\bibitem{C59} A. Clebsch, Ueber die Integration der hydrodynamischen
Gleichungen, \textit{J. reine angew. Math.} \textbf{56 }, 1-10, (1859).

\bibitem{L63} C.C. Lin, Hydrodynamics of Helium II. \textit{Proc. Int. Sch
Phys.} Course XXI, pp. 93-146, New York, Academic, 1963.

\bibitem{R2004} Yu. A. Rylov, Hydrodynamic equations for incompressible
inviscid fluid in terms of generalized stream function. \textit{Int. J.
Math. \& Mat. Sci.} vol. \textbf{2004}, No. 11, 21 February 2004, pp.
541-570. (Available at http://arXiv.org/abs/physics/0303065).

\bibitem{S88} R. Salmon, Hamilton fluid mechanics, \textit{Ann. Rev. Fluid.
Mech.} \textbf{20}, 225-256, {1988}.
\end{thebibliography}
\end{document}